\documentclass[aps,prl,amssymb]{revtex4}

\usepackage{graphicx}
\usepackage{bm}

\def\ep {\epsilon}

\def\e2 {\epsilon-\epsilon_k}
\def\be {\begin{equation}}
\def\ee {\end{equation}}
\def\bea {\begin{eqnarray}}
\def\eea {\end{eqnarray}}
\def\om {\omega}
\def\Om {\Omega}

\def\cd {c^{\dagger}}

\def\ua {\uparrow}
\def\da {\downarrow}
\def\si {\sigma}

\def\dt {\tau-\tau'}
\def\de {\delta}
\def\De {\Delta}
\def\g {\gamma}
\def\th {\theta}
\def\ze {\zeta}
\def\la {\lambda}

\begin{document}

\centerline{Annals of Physics {\bf 349}, 100-116 (2014) - 
	DOI: 10.1016/j.aop.2014.06.015	   }

\title{Variational wavefunction for multi-species spinful 
fermionic superfluids and superconductors }

\author{ George Kastrinakis$^*$}

\affiliation{ Institute of Electronic Structure and Laser (IESL), 
Foundation for Research and Technology - Hellas (FORTH),
P.O. Box 1527, Iraklio, Crete 71110, Greece}

\author{Received 31 March 2014, Accepted 13 June 2014, 
Available online 20 June 2014}

\begin{abstract}

We introduce a new fermionic variational wavefunction,
generalizing the Bardeen-Cooper-Schrieffer (BCS) wavefunction, 
which is suitable for 
interacting multi-species spinful systems and sustaining superfluidity. 
Applications range from quark matter
to the high temperature superconductors.
A wide class of Hamiltonians, comprising interactions and hybridization
of arbitrary momentum dependence 
between different fermion species, can be treated in a comprehensive
manner. This is the case, as both the intra-species and the inter-species
interactions are treated on equally rigorous footing, which is 
accomplished via the introduction of a new quantum index attached to
the fermions. The index is consistent with known fermionic physics,
and allows for heretofore unaccounted fermion-fermion correlations. 
We have derived the finite temperature version of the theory, thus
obtaining the renormalized quasiparticle dispersion relations,
and we discuss the appearance of charge and spin density wave order.

We present numerical solutions for two electron species in 2 dimensions.
Based on these solutions, we show that, for equivalent spin up and down 
fermions, the Fermi occupation factor (per spin) equals 1/2 
deep in the Fermi sea. This constitutes a unique experimental prediction 
of the theory, both for the normal and superfluid states. Interestingly, 
this result, obtained in the thermodynamic limit, is consistent with 
Fermi occupation factor (in-)equalities for finite systems of electrons, 
derived (in a different context)
by Borland and Dennis, J. Phys. B {\bf 5}, 7 (1972) and
by Altunbulak and Klyachko, Commun. Math. Phys. {\bf 282}, 287 (2008). 
\end{abstract}

\maketitle

\vspace{0.3cm}
\centerline{\bf 1. Introduction}
\vspace{0.3cm}

The Bardeen-Cooper-Schrieffer (BCS) wavefunction  
$\Psi_{\text {BCS}}$ \cite{bcs} has set a paradigm 
for the description of fermionic superfluids.
$\Psi_{\text {BCS}}$ is meant to describe systems with a 
single species of spinful fermions. Early on, it was extended to systems with 
{\em two} distinct fermion species by Moskalenko \cite{mosk} and by
Suhl, Matthias, and Walker \cite{suhl}. These approaches can treat 
{\em strictly} BCS-type inter-species and intra-species interactions 
{\em only}.

Despite the appearance of numerous
papers treating {\em multi-species} fermionic systems,
the main challenge in these systems
has remained open ever since.
Namely, how can {\em both} the intra-species and the inter-species
interactions in their {\em totality} be treated on equal footing?
This problem is relevant for many different fermionic systems
such as quark matter \cite{qua}, nuclei \cite{star}, 
neutron stars \cite{star},
superconducting grains \cite{delf}, cold atoms \cite{melo}, 
graphene \cite{graf1,graf2}, APt$_3$P (A=Sr,Ca,La) \cite{pt3p},
and high-temperature superconductors, 
i.e. both copper oxides \cite{eme,gk0,tim} and iron pnictides \cite{john}. 
E.g. in solids, electrons in different bands, with different dispersion
relations and effective masses, correspond to different species.

In this work, we introduce 
a variational wavefunction $\Psi$ for fermionic systems with two or more 
different species of spinful fermions, which can fulfill this purpose - c.f.
the discussion following eqs. (\ref{ham1}), (\ref{ham2}). This is made 
possible through the use of a {\em novel quantum index}, which is 
attached to the fermions and is related to the internal structure of 
the quantum state.
The physical meaning of the index is this. Every fermion of given 
momentum and spin can be considered as participating in an appropriate
superposition of states, which is made possible by the index.
Hence, this index serves to enumerate the {\em disentangled} 
components of the quantum state as a function of both momentum and spin.
This index has {\em no classical} correspondence.

The theoretical motivation for the introduction of the new fermion
index can be explicitly stated. The index allows to consider a multitude
of fermion-fermion correlations in an adequate superposition. This was 
{\em not} possible thus far. These correlations, 
contained in $\Psi$, allow for a {\em comprehensive account
of the generic momentum dependence} of both the intra-species and the
inter-species interactions. This is clearly seen in the expression of the 
expectation value of the total energy $\langle H\rangle$. In this
strong coupling approach, both 
types of interactions are treated on an equally rigorous footing.
For the most part of this paper, we restrict ourselves to pairs of 
particles with opposite momenta. In Section 6 we show
how more involved correlations of particles with non-zero total 
momentum can be treated, yielding, inter-alia, charge and/or spin density 
wave order, {\em irrespectively} of the existence of superconductivity
in the system.

We point out that the theory makes a {\em unique experimental prediction} 
both for the normal and superfluid/superconducting states. Namely,
the Fermi occupation factor (per spin) equals 1/2 deep in
the Fermi sea. This is due to to new quantum index, and it is discussed
in Section 4. It is shown therein that this configuration minimizes
the kinetic energy, and also the total energy of the system, if the 
interactions are not very strong.

In the foregoing we will restrict ourselves to the case of {\em two}
fermion species, which is sufficient in order to demonstrate the 
features of the whole theory involving the new $\Psi$. 
It is straightforward to generalize the formalism
to more than two fermion species.

The relevance of the BCS states to the calculation of $T_c$ for the
multilayer copper oxide superconductors
\cite{gk0} provided a motivation for this work, in an effort to consider 
relevant pairing correlations, beyond the standard BCS ones.

This paper is organized as follows.
In Section 2 we introduce the new wavefunction $\Psi$.
In Section 3 we show how relevant algebraic calculations proceed,
including the expectation value of the energy.
In Section 4 we discuss the ground state of the theory,
and we provide explicit such numerical solutions for a system 
of two electron species (bands) in 2 dimensions.
In Section 5 we discuss the finite temperature dependence of the theory,
from which the quasiparticle dispersion relations 
and the critical temperature $T_c$ emanate.
In Section 6 we discuss the appearance of charge and spin density 
wave order. We summarize in Section 7. There are also four
Appendices. In Appendices A and B 
we present two different spin triplet versions
of the new wavefunction $\Psi$. In Appendix C we discuss the main energy
minimization conditions. In Appendix D we present the complete derivation of
the finite temperature dependence of the theory.

\vspace{0.6cm}
\centerline{\bf 2. The new wavefunction $\Psi$ }
\vspace{0.3cm}

For reference, $|\Psi_{\text {BCS}}\rangle  = \prod_{k} 
(u_{k} + v_{k} \; \cd_{k,\ua} \cd_{-k,\da})|0\rangle $ \cite{bcs},
with the creation/annihilation operators $\cd_{k,\si}/c_{k,\si}$ describing
fermions with momentum $k$ and spin $\si$, and $|0\rangle $ being the vacuum
state.
Now let the usual fermionic operators be $\cd_x/c_x$ with $x=\{i,k,\si\}$, 
where $i$ denotes the fermion {\em species}.
Thereby we introduce the {\em new fermionic operators}
$\cd_{x,\nu}/c_{x,\nu}$, with the additional new index denoted as $\nu,\mu$,
obeying the anticommutators ($\{a,b\}=ab+ba$)

\be
\{c_{x,\mu},c_{y,\nu}\}=0 \;\;,\;\;
\{c_{x,\mu},\cd_{y,\nu}\}=\de_{xy} \; \de_{\mu\nu}  \;\;, 
\ee

and we write the usual  $\cd_x/c_x$ as the superposition

\be
\cd_x = \sum_{\nu=1}^{N_o} \g_{x,\nu}^* \; \cd_{x,\nu} \;\; , \;\;
c_x = \sum_{\nu=1}^{N_o} \g_{x,\nu} \; c_{x,\nu} \;\; .  \label{anac}
\ee
$N_o$ is discussed below.
The usual anticommutation relations of $\cd_x/c_x$ are preserved by imposing 
the normalization condition
\be
 \sum_{\nu=1}^{N_o} |\g_{x,\nu}|^2 = 1 \;\;, \;\; \label{norma}
\ee
for the complex weight coefficients $\g_{x,\delta}$, which are to be 
determined via the energy minimization procedure
below, and the solution of 
equations (\ref{eqm1})-(\ref{eqm2}), while

\be
\{c_{x},c_{y,\nu}\}=0 \;\;,\;\;
\{c_{x},\cd_{y,\nu}\}=\de_{xy} \; \g_{x,\nu} \;\;.
\ee
Eq. (\ref{norma}) simply means that for every single fermion 
the components with which it participates in a superposition of states
add up to precisely {\em one fermion} (nothing less or more).

Considering two species of fermions, we also introduce
\be
A_{i,k,\nu}^{\dagger} = u_{i,k} +
v_{i,k} \; \cd_{i,k,\ua,\nu} \; \cd_{i,-k,\da,\nu}
+ s_{i,k,\ua} \; \cd_{i,k,\ua,\nu} \; \cd_{j,-k,\da,\nu}
+ s_{i,k,\da} \; \cd_{i,-k,\da,\nu} \; \cd_{j,k,\ua,\nu}
\;\;.  \label{eqada}
\ee
$A_{i,k,\nu}^{\dagger}$ is a bosonic operator, creating spin singlet
pairs of fermions (for triplet pairs c.f. eq. (\ref{eqtr}) below),
and $(i,j)=\{(1,2),(2,1)\}$.

We form the following multiplet of $A_{i,k,\nu}^{\dagger}$'s
\be
M_k^{\dagger}  = A_{1,k,\nu=1}^{\dagger} \; A_{1,-k,\nu=1}^{\dagger} \;
A_{2,k,\nu=2}^{\dagger} \; A_{2,-k,\nu=2}^{\dagger} \;\;. \label{eqmm}
\ee
$M_k^{\dagger}$ creates all relevant states with momenta $\pm k$. 
We consider the most simple case for the new index, i.e. taking only two 
discrete values, say 1 and 2, with 
$N_o=2$. As mentioned after eq. (\ref{eqap}) below, the index can be
{\em continuous}, in principle.

The new index allows for the bookkeeping
of a superposition of states of a given particle,
i.e. same $x=\{i,k,\si \}$, without the difficulties due to entanglement
within the multiplet $M_k^{\dagger}$, if the index were removed.
In that case, the treatment of the coherence factors 
$u_{i,k},v_{i,k},s_{i,k,\si}$ 
is prohibitively complicated, especially in the thermodynamic limit.

We note that there is {\em no change} whatsoever implied in the Hamiltonian 
or 
in the representation of any observable, as a result of the introduction
of the new $\cd_{x,\nu},c_{x,\nu}$'s. The new index is consistent with 
known fermionic physics.

Now we introduce the {\em disentangled} state
\be
|\Psi\rangle  = \prod_{k'} M_k^{\dagger} \; |0\rangle  \;\;,
\ee
where the prime implies that $k$ runs over {\em half} the momentum space.
Note that {\em all} $A_{i,k,\nu}^{\dagger}$'s in $|\Psi\rangle$ {\em commute 
with each other.}

$\Psi$ generalizes $\Psi_{\text {BCS}}$ and
sustains superfluidity. Spin triplet versions of $\Psi$ can be found  
in Appendices A and B.
This wavefunction makes particularly sense for two or more fermion 
species, with an interaction between different species. It can obviously
be generalized for three or more fermion species. Moreover, a related 
wavefunction for a {\em single} fermion species system can be written. 
In this case the new quantum index becomes relevant in the limit of strong 
interaction, and it allows to consider correlations between 2 and 4
fermions with different momenta \cite{gk2}.
Further, a wavefunction of this type using the new quantum index can 
be written 
in the real space representation instead of the momentum space one.
$\Psi$ opens up a very promising avenue for the treatment
of many-body systems, as can be seen from the discussion which follows.

$\Psi$ allows for inequivalence between spin up and down fermions.  
Plus, it allows for the comprehensive variational treatment
of a  wider class of Hamiltonians than sheer 
BCS type, e.g. comprising interactions and hybridization of arbitrary 
momentum dependence
between different fermion species, similar to the well 
known manner of the BCS-Gorkov theory \cite{bcs},\cite{agd}.

The normalization condition $\langle \Psi|\Psi\rangle =1$ implies
\be
u_{i,k}^2+|v_{i,k}|^2+|s_{i,k,\ua}|^2+|s_{i,k,\da}|^2=1\; , \;
0\leq u_{i,k}^2,|v_{i,k}|^2,|s_{i,k,\ua}|^2,|s_{i,k,\da}|^2 \leq 1 \;\;,
\ee
thus allowing to treat these coherence factors as
\bea
u_{i,k}=\cos(\theta_{i,k}) \cos(\phi_{i,k}) \;\; , \;\;
v_{i,k}=\sin(\theta_{i,k}) \cos(\phi_{i,k}) \exp(i a_{i,k}),   \nonumber \\
s_{i,k,\ua}=\cos(\delta_{i,k}) \sin(\phi_{i,k}) \exp(i b_{i,k,\ua})\;\; ,
\;\; s_{i,k,\da}=\sin(\delta_{i,k}) \sin(\phi_{i,k}) \exp(i b_{i,k,\da}) . 
\eea

\vspace{0.6cm}
\centerline{\bf 3. Algebraic calculations with $\Psi$ }
\vspace{0.3cm}

{Algebraic calculations with $\Psi$ are straightforward}. 
Below we elaborate on the case of two fermion species with dispersions
$\ep_{i,k,\si}=\ep_{i,-k,\si}$.
We have
\bea
c_{1,k,\ua} M_k^{\dagger} |0\rangle = 
\g_{1k \ua ,1}
(v_{1,k} \cd_{1,-k,\da,\nu=1} +  s_{1,k,\ua} \cd_{2,-k,\da,\nu=1})
 A_{1,-k,\nu=1}^{\dagger} 
A_{2,k,\nu=2}^{\dagger} A_{2,-k,\nu=2}^{\dagger} |0\rangle \nonumber \\
- \g_{1k \ua ,2} \; s_{2,k,\da} \;  
\cd_{2,-k,\da,\nu=2} \; A_{1,k,\nu=1}^{\dagger} \; 
A_{1,-k,\nu=1}^{\dagger} \; A_{2,-k,\nu=2}^{\dagger} \; |0\rangle \;\;,
\eea
and
\bea
\langle 0| \; M_k \; \cd_{1,k,\ua} c_{1,k,\ua} \; M_k^{\dagger} \; |0\rangle =
|\g_{1k \ua ,1}|^2 \; (|v_{1,k}|^2 + |s_{1,k,\ua}|^2) 
+  | \g_{1k \ua ,2} \; s_{2,k,\da}|^2 \;\;.
\eea
Further,
\bea
c_{2,k,\ua} \; M_k^{\dagger} \; |0\rangle = 
\g_{2k \ua ,2} 
(v_{2,k} \; \cd_{2,-k,\da,\nu=2} +  s_{2,k,\ua} \; \cd_{1,-k,\da,\nu=2})
\; A_{1,k,\nu=1}^{\dagger} \;
A_{1,-k,\nu=1}^{\dagger} \; A_{2,-k,\nu=2}^{\dagger} \; |0\rangle \nonumber \\
- \g_{2k \ua ,1} \;
s_{1,k,\da} \; \cd_{1,-k,\da,\nu=1} \; A_{1,-k,\nu=1}^{\dagger} \; 
A_{2,k,\nu=2}^{\dagger} \; A_{2,-k,\nu=2}^{\dagger} \; |0\rangle \;\;,
\eea
which yields
\be
\langle 0| \; M_k \; \cd_{2,k,\ua} c_{1,k,\ua} \; M_k^{\dagger} \; |0\rangle =
  -( \g_{1k \ua ,1} \; \g_{2k \ua ,1}^* \; v_{1,k} \;s_{1,k,\da}^* 
+ \g_{2k \ua ,2}^* \; \g_{1k \ua ,2} \; v_{2,k}^* \;s_{2,k,\da} ) \;\;.\;\;
\ee
Moreover,
\bea
 c_{2,-k,\da}\; c_{1,k,\ua} \; M_k^{\dagger} \; |0\rangle = 
\g_{2 -k \da ,1} \; \g_{1 k \ua ,1} \; s_{1,k,\ua} \; A_{1,-k,\nu=1}^{\dagger} 
\; A_{2,k,\nu=2}^{\dagger} \; A_{2,-k,\nu=2}^{\dagger} \; |0\rangle 
\nonumber \\
- \g_{2 -k \da ,2} \; \g_{1k \ua ,2} \; 
s_{2,k,\da} \; A_{1,k,\nu=1}^{\dagger} \; 
A_{1,-k,\nu=1}^{\dagger} \; A_{2,-k,\nu=2}^{\dagger} \; |0\rangle \;\;,
\eea
and
\be
\langle 0| \; M_k \; c_{2,-k,\da} c_{1,k,\ua} \; M_k^{\dagger} \; |0\rangle =
 \g_{1k \da ,1} \; \g_{1k \ua ,1} \; u_{1,k} \;s_{1,k,\ua}-
\g_{1k \da ,2} \; \g_{1k \ua ,2} \; u_{2,k} \;s_{2,k,\da}  \;\;.\;\;
\ee
Also
\be
\langle 0| \; M_k \; c_{1,-k,\da} c_{1,k,\ua} \; M_k^{\dagger} \; |0\rangle =
 \g_{1k \da ,1} \; \g_{1k \ua ,1} \; u_{1,k} \;v_{1,k} \;\;.\;\;
\ee

Likewise, and using the commutativity of $A_{i,k,\nu}^{\dagger}$'s, we obtain 
($\langle B \rangle =\langle \Psi | B | \Psi \rangle)$ 
\bea
n_{i,k,\si}  = \langle \cd_{i,k,\si} c_{i,k,\si} \rangle
= |\g_{i k \si ,i}|^2 \; (|v_{i,k}|^2 + |s_{i,k,\si}|^2) +  
|\g_{i k \si ,j} \; s_{j,k,-\si}|^2
\;\;, \;\; (i,j)=(1,2),(2,1) \;\; , \label{mel1} \\
z_{k,\si} = \langle \cd_{2,k,\si} c_{1,k,\si} \rangle =
-\text{sgn}(\si) 
\;(\g_{1 k \si,1}\; \g_{2 k\si,1}^*\; v_{1,k} \;s_{1,k,-\si}^* 
+ \g_{2 k \si,2}^* \; \g_{1 k\si,2}\; v_{2,k}^* \;s_{2,k,-\si})
\; \; ,  \\
g_{k,\si} = \langle c_{2,-k,-\si} c_{1,k,\si} \rangle
= ( \g_{1 k \si,1}\; \g_{2 k -\si,1}\; u_{1,k} \;s_{1,k,\si}
- \g_{2 k -\si,2}\; \g_{1 k\si,2}\; u_{2,k} \;s_{2,k,-\si} )  
\; \; ,  \\
f_{i,k,\si} = \langle c_{i,-k,-\si} c_{i,k,\si} \rangle
= \g_{i k \si ,i} \; \g_{i k -\si ,i} \; u_{i,k} \; v_{i,k}  \label{mel2}
 \;\; .
\eea

A general Hamiltonian for two fermion species interacting via intra-species 
potentials $V_{1,2}$ and via an inter-species potential $F_q$, 
and hybridizing via $h_k$, is
\bea
H = \sum_{i,k,\si} \xi_{i,k,\si} \;  \; \cd_{i,k,\si} c_{i,k,\si}
+ \sum_{k,\si} h_k  
\left( \cd_{1,k,\si} c_{2,k,\si} + \cd_{2,k,\si} c_{1,k,\si} \right)
\label{ham1}  \\
+ \frac{1}{2} \sum_{i,k,p,q,\si,\si'} V_{i,q} \; 
\cd_{i,k+q,\si} \cd_{i,p-q,\si'} c_{i,p,\si'}c_{i,k,\si}
+ \sum_{k,p,q,\si,\si'} F_q \; \cd_{1,k+q,\si} \cd_{2,p-q,\si'} 
c_{2,p,\si'}c_{1,k,\si}
\;\;,  \nonumber 
\eea
with $i=1,2$, $\xi_{i,k,\si}=\ep_{i,k,\si}-\mu_{i,\si}$ and 
$\mu_{i,\si}$ the chemical potential.
Note that both $V_{i,q}$ and $F_q$ are taken to have a {\em generic momentum 
dependence}. We do {\em not} restrict ourselves to some kind of 
separable potentials
or, otherwise, very special type of potentials.
Here, the usual BCS pairing potential is just the sub-term
$\sum_{i,k,p}V_{i,k-p}\; \cd_{i,k,\ua} \cd_{i,-k,\da} 
c_{i,-p,\da}c_{i,p,\ua} $ of the single species potential.

Considering $\Psi$ and eqs. (\ref{mel1})-(\ref{mel2}) above, we
evaluate $\langle H\rangle =\langle \Psi|H|\Psi\rangle $.
Then
\bea
\langle H\rangle  = \sum_{i,k,\si} \xi_{i,k,\si} \; n_{i,k,\si} 
+ \sum_{k,\si}  \; h_k \;( z_{k,\si} +  z_{k,\si}^* ) 
+\frac{1}{2} \sum_{i,k,p,\si} \left( V_{i,q=0}\; - V_{i,k-p} \right)
\; n_{i,k,\si} \;n_{i,p,\si}
\label{ham2} \\
+ \frac{1}{2} \sum_{i,k,p,\si} V_{i,k-p}\; f_{i,k,\si} \; f_{i,p,\si}^*
-\sum_{k,p,\si} F_{k-p} \; z_{k,\si} \; z_{p,\si}^* 
+  F_{q=0}  \; n_{1} \;n_{2}
+ \sum_{k,p,\si} F_{k-p} \; g_{k,\si} \; g_{p,\si}^*
\nonumber \;\;,  
\eea
with $(i,j)=\{(1,2),(2,1)\}$ and
the total filling factor per species is $n_i=\sum_{k,\si} n_{i,k,\si}$.  
The various terms of $\langle H\rangle $ are derived by exhausting all 
possible combinations of expectation values of two and four fermion creation
and annihilation operators. Due to the specific form of $|\Psi\rangle $ 
considered, the above expression for $\langle H\rangle $ coincides with 
the one given by the Hartree-Fock-Bogoliubov approximation
(within which  the expectation value of products of 4 operators equals
$\langle c_1 c_2 c_3 c_4 \rangle=\langle c_1 c_2 \rangle  
\langle c_3 c_4 \rangle
- \langle c_1 c_3 \rangle  \langle c_2 c_4 \rangle
+ \langle c_1 c_4 \rangle  \langle c_2 c_3 \rangle$).

The first term in the second line is exactly
the usual BCS pairing term, and the last term is the equivalent inter-species
pairing term due to $F_q$. Manifestly $\langle H\rangle$ takes into account 
the potentials $V_{i,q}$ and $F_q$ in their {\em entirety and not in some
partial manner} - as the case is with the BCS treatment \cite{bcs,mosk,suhl}.
Of course, this is a {\em strong coupling} approach (BCS, in contrast, omits
terms such as $\sum_{i,k,p,\si} \left( V_{i,q=0}\; - V_{i,k-p} \right)
\; n_{i,k,\si} \;n_{i,p,\si}$, and
$\sum_{k,p,\si} F_{k-p} \; z_{k,\si} \; z_{p,\si}^*$).
Actually, expanding the Hilbert space spanned by $ |\Psi\rangle $, 
by including additional 2-fermion correlations, as e.g. in 
eqs. (\ref{eqap}),(\ref{eqtr}), yields additional terms in $\langle H\rangle$,
which depend on $V_{i,q}$ and $F_q$. In principle, this procedure yields
even lower estimates for the ground state energy.

The simplest case to consider is with the up and down
spins being equivalent, i.e. with $|\sin(\de_{i,k})|=|\cos(\de_{i,k})|=
1/\sqrt{2}$ and $\xi_{i,k,\si}=\xi_{i,k}$. We also make the choice 
\be
\g_{i k, i} = \cos(\eta_{i,k}) \;\;, \;\; 
\g_{i k, j} =\sin(\eta_{i,k}) \; \exp(i \om_{i,k})  \;\; ,
\ee 
which is justified in the discussion after eq. (\ref{gapb}), 
and $(i,j)=\{(1,2),(2,1)\}$. To obtain the ground states we minimize 
$E = \langle H\rangle $ with respect to the 
angles $\theta_{i,k}, \phi_{i,k}, a_{i,k}, b_{i,k}$, $\om_{i,k}$ and 
$\eta_{i,k}$ 
\be
0= \frac{\partial E}{\partial \theta_{i,k}}=
\frac{\partial E}{\partial \phi_{i,k}}
= \frac{\partial E}{\partial a_{i,k}} = \frac{\partial E}{\partial b_{i,k}}
= \frac{\partial E}{\partial \om_{i,k}}=\frac{\partial E}{\partial\eta_{i,k}}
\;\; .\;\; \label{eqm1}
\ee

We elaborate on the minimization conditions (\ref{eqm1}) in Appendix C.

Focusing on the condition for $\eta_{i,k}$ we have
\be
0= \frac{\partial E}{\partial \eta_{i,k}}  \;\; \rightarrow  \;\; 
 \eta_{i,k} = \arctan (N_{i,k}/D_{i,k})  \;\; .  \label{eqm2} 
\ee
Here 
\bea
N_{i,k} =  |s_{j,k,\si}|^2 \; \sin(\eta_{i,k}) \; \Xi_{i,k}
+ \g_{j k ,j} \; \text{Re} \{ T_{j,k}  \}\;\;,  \;\;
\Xi_{i,k}=\xi_{i,k} +\sum_{p} \left( V_{i,q=0} 
- V_{i,k-p} \right) n_{i,p,\si} 
+ F_{q=0} \; n_{j}  \;\;, \nonumber \\     \label{eqom}
D_{i,k} =\g_{i k ,i} [ (|v_{i,k}|^2 + |s_{i,k}|^2) \; \Xi_{i,k}  
-  \text{Re} \{ \Delta^{*}_{i,k} \; u_{i,k} v_{i,k} \} ] 
+  \sin(\eta_{j,k}) \; \text{Re} \{ T_{i,k} \} \;\;, \;\; \\  
T_{i,k}= (h_k - S_k^* )  \; \sin(\th_{i,k}) \; \sin(2 \phi_{i,k})
\; \exp(i \; (-1)^i \; \Omega_{ij,k}) /(2\sqrt{2})
+ u_{i,k} \; s_{i,k,\si}\; \Phi_k^* \; \exp(i \om_{j,k})\;\;, \;\;   
\nonumber \\ 
S_k =\sum_{p} F_{k-p} \; z_{p,\si}  \;\;, \;\; 
\Phi_k =\sum_{p} F_{k-p} \; g_{p,\si}  \;\;, \;\;  
 \Om_{ij,k} = b_{i,k}-a_{i,k}+\om_{j,k}  \;\;. \nonumber
\eea
(Note that $i$ stands both for the index $i=1,2$ and for the imaginary
$i^2=-1$, the latter appearing in the argument of the exponential function.) 
The generalization of the BCS gap is
\bea
 \Delta_{i,k} = - \sum_p V_{i,k-p} \;  f_{i,p,\si} \;\; . \;\; \label{gapb}
\eea
We see that for $F_q \rightarrow 0$ and $h_k \rightarrow 0$ the angles
$\eta_{i,k}$ go {\em smoothly} to zero. In this case only 
$\g_{i k ,i} \rightarrow 1$ survive, and $\g_{i k ,j} \rightarrow 0$.
That is, only one term of the superposition in eq. (\ref{anac}) 
survives, consistent with the "conventional" case.

\vspace{0.6cm}
\centerline{\bf 4. The ground state of the theory}
\vspace{0.3cm}

Equations (\ref{eqm1}) are necessarily satisfied by the 
ground state. However, they should be supplemented by additional
conditions, which specify in a {\em unique manner} the ground state.
Overall, this constitutes a {\em highly non-trivial} and non-convex
optimization problem,
which is difficult to solve. C.f. below.

\begin{table}\centering

\begin{tabular}{|c|c|c|c|c|} \hline
  
$ V_0 $ & $E(V_1=V_2=5)$ & $ \De_1,\; \De_2 $ & $ E(V_1=V_2=10)$ & 
$ \De_1,\; \De_2 $ \\
\hline

$ 0.5 $ & $ -3.711 $ & $ d,\;0$ & $ -3.311$ & $ 0,\;d$ \\
\hline

$ 1 $ & $ -4.815$ & $ d,\;0$ & $ -4.268 $ & $ d,\;0$ \\
\hline

$ 2 $ & $ -6.768 $ & $ d,\;0$ & $ -6.317 $ & $ d,\;0$ \\
\hline

$ 3 $ & $ -8.540 $ & $ d,\;0$ & $ -8.216 $ & $ d,\;0$ \\
\hline

$ 4 $ & $ -10.841 $ & $ d,\;0$ & $ -10.354 $ & $ d,\;0$ \\

\hline

\end{tabular}

{ Table 1. Ground state energies for the parameters shown. Also shown
the symmetry of the respective superconducting gaps $\De_{1,2}$, 
with $d$ standing for
$d_{x^2-y^2}$-wave and 0 for absence of a gap. C.f. text. } 

\end{table}

We thus adopted the following 
procedure in order to locate the ground state. We solve numerically
equations (\ref{eqm1}) by (fully deterministic)
iteration. We implement an exhaustive search in the space of initial
conditions of the solutions and in the space of certain control
parameters of the (custom made) algorithm used. In the end, among all 
solutions of equations (\ref{eqm1}) obtained, we select the state 
with the minimum energy $E=\langle H\rangle $ as the ground state.

We present self-consistent numerical solutions for a system composed of 
two different species (bands) of electrons in 2 dimensions. 
We use an $N \times N$ discretization of the Brillouin zone, with $N=120$.
Overall, we have 2 (fermion species) $ \times $ 6 
(different variables/angles per fermion) $ \times $ $ N^2$, divided by
4 (due to the $C_4$ symmetry of the Brillouin zone), amounting
to a total of 43,200 variables. 

For our numerical {\em examples}, we use realistic tight-binding dispersion 
relations
and realistic effective intra-species and inter-species potentials.
We consider
$\ep_{i,k} = -2 t_i (\cos{k_x}+\cos{k_y}) 
- 4 t_i'\cos{k_x} \cos{k_y}-2 t_i'' (\cos{2k_x}+\cos{2k_y}) $, 
with the momentum $k=(k_x,k_y), \; k_x,k_y=[-\pi,\pi]$,
and $t_i=1$, $t_i'=-0.35$, $t_i''=0.12$. The hybridization $h_k=0$. 
The filling factors are $n_1=0.91$
and $n_2=0.81$ - and these correspond to different chemical potentials,
which are calculated self-consistently.
For the intra-species
potential we consider 
\be
V_{i,q}=V_{i} \sin^2(q_x/2) \sin^2(q_y/2 )  \;\;,\;\;
\ee
which is peaked at $ Q =(\pm \pi,\pm \pi)$. For the inter-species
potential we consider 
\be
F_q = V_0 [\cos(q_x/2) + \cos(q_y/2)]  \;\;.\;\;
\ee
All energies are measured in units of $t_1$.
In table 1 we show the ground state energy,  
as a function of $V_1=V_2$ and $V_0$, and the gap symmetry. These states 
$\Psi$ have $d_{x^2-y^2}$-wave superconducting gaps (the gap symmetry 
is due to the $V_{i,q}$ used \cite{gk0}), for moderate values of $V_0$,
as shown in the table. 

\begin{figure}[ht]
\includegraphics[height=7cm]{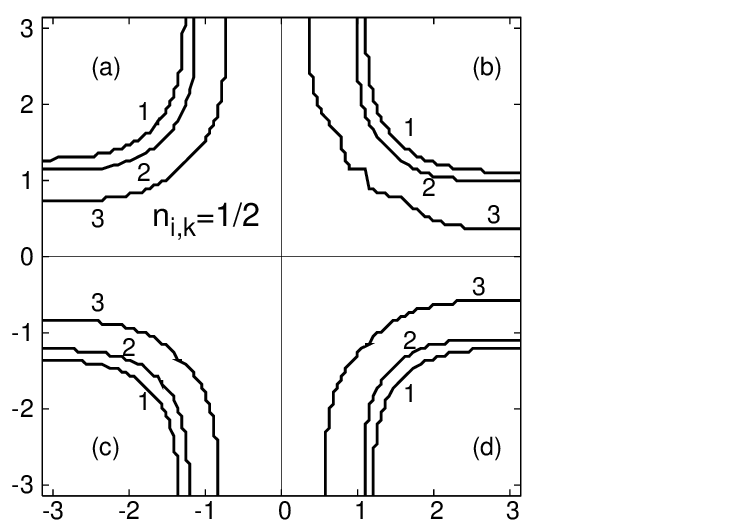}
\caption{ The occupation factor $n_{i,k,\si}$ of the ground state
and the Fermi surfaces as a function of momentum $k=(k_x,k_y)$ 
in the Brillouin zone, as obtained numerically.
These are 4-fold symmetric
(group $C_4$) in momentum space, and figs. (a)-(d) each display 
1/4 of the Brillouin zone. In figs. (a),(b) $V_1=V_2=5$. 
$V_o=0.5$ for fig. (a) and $V_o=3$ for fig. (b).  
In figs. (c),(d) $V_1=V_2=10$. 
$V_o=0.5$ for fig. (c) and $V_o=3$ for fig. (d). 
In each quadrant, the lines marked 1 and 2 are the
Fermi surfaces for species 1 and 2, respectively. 
Inside line 3 is the locus of momenta
with $n_{i,k,\si}=1/2$. Between line 3 and line 1 $n_{1,k,\si}=1$, and 
between line 3 and line 2 $n_{2,k,\si}=1$. $n_{i,k,\si}=0$ above the 
respective Fermi surfaces. C.f. text. Also, some of the solutions
have $n_{i,k,\si}=1/2$ along the diagonals even within the outer
momentum shell, where otherwise $n_{i,k,\si}=1$.    }
\end{figure}

We demonstrate a {\em novel feature of the ground state} at zero 
temperature. Namely, for a broad range of the inter-species potential $F_q$, 
the Fermi occupation factor $n_{i,k,\si}$
equals 1/2 for a symmetric locus of momenta around zero 
momentum, as the angles $|\eta_{i,k}|=\pi/2-\de \eta_{i,k}$ and
$|\phi_{i,k}|=\pi/2-\de \phi_{i,k}$ therein
- c.f. eq. (\ref{mel1}). Both $|\de \phi_{i,k}|, \; |\de \eta_{i,k}|$
are very small - see eqs. (\ref{def}), (\ref{dee}) and the discussion 
following them in Appendix C. 
The factor 1/2 simply reflects
the equivalence between up and down spin species.
For higher momenta, and up to the Fermi momentum, $n_{i,k,\si}$ is 
equal to 1. C.f. fig. 1. 
We note that this inner locus is the {\em same} for both electron species. 

{\em This $n_{i,k,\si}=1/2$ configuration is favored by kinetic energy 
minimization.} Consider a non-interacting 1-band model, with chemical 
potential $\mu=E_F$. Suppose that $n_{k,\si}=1/2$ for $\ep_k \leq E_o$
and $n_{k,\si}=1$ for $E_o < \ep_k \leq E_F$. Then $E$ from eq. 
(\ref{ham2}) is $E=\sum_{k,\si} \; \xi_{k,\si} \; n_{k,\si} = 
-N_F \; (E_F^2+E_o^2/2-E_F E_o)$, 
where, for simplicity, a constant density of states $N_F$
is assumed. Now consider the same system but with the conventional 
$n_{k,\si}'=1$ for all $\ep_k \leq E_F'=\mu'<\mu$, 
yielding $E'= -N_F \; (E_F')^2$.
We see that $E < E'$ if $(E_F')^2 < E_F^2+E_o^2/2-E_F E_o$.
For a broad range of band fillings this inequality can be satisfied,
resulting in the unusual 1/2 occupancy.

As shown in fig. 1, this configuration survives for finite 
positive interactions $V_{i,q},F_q>0$,
provided that $F_q$ is {\em not} very strong. In the latter case, the unusual
occupancy equal to 1/2 disappears gradually from the core of the Fermi
sea.

Finite hybridization $h_k \neq 0$ does not modify this picture.

The matter of constraints on the Pauli principle for discrete
systems with a {\em finite} number of electrons has been discussed
in the literature \cite{db,kly,chris} (and therein). In this context,
(in-)equalities involving the expectation value of Fermi occupation factors 
$\la_i$ for different single particle states, labeled by $i$, have been
derived. It is noteworthy that setting $\la_i$ equal to $n_{k,\si}=1/2$
satisfies (in-)equalities (2) and (3) in \cite{chris}, which were
actually first derived in \cite{db}. Further, $n_{k,\si}=1/2$ satisfies
e.g. inequalities (4) in \cite{kly}. Of course, our result $n_{k,\si}=1/2$
has been derived in a totally different context, i.e. for a many-body
system in the presence of the new quantum index. One could speculate
that a connection between the two kinds of systems exists, the exact
nature of which is not clear at present.

\vspace{0.6cm}
\centerline{\bf 5. The finite temperature dependence of the theory:  }
\centerline{\bf quasiparticle dispersion relations and critical 
temperature $T_c$ }
\vspace{0.2cm}

The finite temperature dependence of the theory has been derived 
through the equations of motion formalism for the Green's functions. 
Full details can be found in Appendix D.
In that frame, both the critical transition temperature $T_c$ into the 
superfluid/superconducting state
and the effective quasiparticle dispersion relations can be calculated. 
For $\xi_{i,k,-\si}=\xi_{i,k,\si}=\xi_{i,k}$ we obtain four different
quasi-particle energy branches - c.f. eq. (\ref{eqdis}) in Appendix D. 
The excited states of this theory are 
not straightforward to obtain, hence we opted for this formalism in 
order to calculate unambiguously the quasi-particle dispersion.

This is simplified for the new ground states thus far 
obtained numerically, as the angles $\eta_{i,k}$ 
and $\phi_{i,k}$
take exclusively the values $|\eta_{i,k}|=0, \pi/2-\de \eta_{i,k}$ and
$|\phi_{i,k}|=0, \pi/2-\de \phi_{i,k}$, with both 
$|\de \phi_{i,k}|, \; |\de \eta_{i,k}|$ being very 
small - see eqs. (\ref{def}), (\ref{dee}) and the discussion 
following them in Appendix C. 
Taking also the hybridization $h_k$=0, we obtain
two branches for the quasiparticle dispersion relation
\be
E_{i,k}^2= \Xi_{i,k}^2+|\Delta_{i,k}|^2 \;\; ,\;\; 
\label{dispe}
\ee
for $(i,j)=\{(1,2),(2,1)\}$ and $\Xi_{i,k}$ is given in eq. (\ref{eqom}).
This dispersion relation is the same as the classic BCS relation 
$E_k^2=\xi_k^2+\Delta_k^2$, modulo the dispersion renormalization 
factor $\Xi_{i,k}-\xi_{i,k}$. We note that (for $n_{i,k,\si}=n_{i,k,-\si}$)
\be
\Xi_{i,k}= \frac{ \partial E}{\partial n_{i,k,\si}} \;\; . \;\;
\ee

The critical temperature $T_c$ is implicitly determined in this theory.
It is the temperature below which the anomalous propagators 
$F_{i,\si}^{\dagger}, F_{ij,\si}^{\dagger}$ - c.f. 
eqs. (\ref{anf1}), (\ref{anf2}) - become {\em non-zero} (also c.f. 
eq. (\ref{eqftc}) and the discussion following it).

\vspace{0.6cm}
\centerline{\bf 6. Charge and spin density wave order }
\vspace{0.2cm}

Charge and spin density wave (CDW/SDW) order can appear 
in a natural manner,
via a simple extension of $\Psi$. Namely, by allowing the total momentum 
of pairs to be finite, which is expected to be favored by the 
finite interspecies potential $F_q$.
For example, we may consider an operator depending on two different
momenta
\bea
A_{i,k,p,\nu}^{\dagger} = u_{i,k} 
+ v_{i,k} \; \cd_{i,k,\ua,\nu}  \; \cd_{i,-k,\da,\nu}
+ v_{i,k,p,\ua} \; \cd_{i,k,\ua,\nu}  \; \cd_{i,p,\da,\nu}  
+ v_{i,-k,-p,\da} \; \cd_{i,-k,\da,\nu}  \; \cd_{i,-p,\ua,\nu}  
\label{eqap}   \\
+ s_{i,k,\ua} \; \cd_{i,k,\ua,\nu} \; \cd_{j,-k,\da,\nu}
+ s_{i,k,\da} \; \cd_{i,-k,\da,\nu} \; \cd_{j,k,\ua,\nu}
+ s_{i,k,p,\ua} \; \cd_{i,k,\ua,\nu} \; \cd_{j,p,\da,\nu}
+ s_{i,k,p,\da} \; \cd_{i,-k,\da,\nu} \; \cd_{j,-p,\ua,\nu}
\;\;.   \nonumber 
\eea
Note that new coefficients $v$ and $s$ are introduced here, which
depend on 2 different momenta $k$ and $p$.
In this case $N_o=4(=2+2)$ - c.f. eq. (\ref{anac}). 
In principle, the new index could be a continuous 
variable, instead of an integer, if a continuous range of momenta would
be correlated with a given $k$. 
This would {\em seem} to be the actual physical case.

For a single pair of such correlated momenta $(k,p)$ we 
form the following multiplet of $A_{i,k,p,\nu}^{\dagger}$'s
\be
M_{k,p}^{\dagger} = A_{1,k,p,\nu=1}^{\dagger} \; A_{1,-k,-p,\nu=2}^{\dagger}\;
A_{1,p,k,\nu=3}^{\dagger} \; A_{1,-p,-k,\nu=4}^{\dagger}\;
A_{2,k,p,\nu=2}^{\dagger} \; A_{2,-k,-p,\nu=1}^{\dagger}\;
A_{2,p,k,\nu=4}^{\dagger} \; A_{2,-p,-k,\nu=3}^{\dagger}\;  \;\; , \;\;
\label{coma}
\ee
which creates all relevant states with momenta $\pm k$,$\pm p$. Note
the {\em particular} assignment of the new index $\nu$, ensuring the 
commutativity of $A_{i,k,p,\nu}^{\dagger}$'s in eq. (\ref{coma}).
Then
$|\Psi\rangle$ is written as
\be
|\Psi\rangle  = \prod_{q' \neq \pm k,\pm p} 
M_q^{\dagger} \; M_{k,p}^{\dagger} \; |0\rangle  \;\;,  \label{eqpsi}
\ee
where the prime implies that $q$ runs over {\em half} the momentum space.
Using eq. (\ref{eqpsi}), we obtain non-zero matrix elements 
$\langle  \cd_{i,-k,\si} c_{i,p,\si} \rangle$,
$\langle \cd_{j,-k,\si} c_{i,p,\si} \rangle$, 
which enter in CDW/SDW. 
That is
\bea
\langle   \cd_{1,-k,\si} \; c_{1,p,\si}   \rangle|_{\si=\ua} =
-\text{sgn}(\si) \big\{ \g_{1,-k,\si,\nu=2}^* \; \g_{1,p,\si,\nu=2}
\; v_{1,-k}^* \; v_{1,-k,-p,-\si}
+ \g_{1,-k,\si,\nu=3}^* \; \g_{1,p,\si,\nu=3} \; v_{1,p} \; v_{1,p,k,-\si}^*
 \big\}\nonumber \\
+\; \g_{1,-k,\si,\nu=1}^* \; \g_{1,p,\si,\nu=1} \;
s_{2,-k,-\si}^* \; s_{2,-k,-p,-\si}
+ \g_{1,-k,\si,\nu=4}^* \; \g_{1,p,\si,\nu=4} \;
s_{2,p,-\si} \; s_{2,p,k,-\si}^*, 
\;\;, \;\;  \label{cdw1}
\eea
\bea
\langle   \cd_{1,-k,\si} \; c_{1,p,\si}   \rangle|_{\si=\da} =
-\text{sgn}(\si) \big\{ \g_{1,-k,\si,\nu=1}^* \; \g_{1,p,\si,\nu=1}
\; v_{1,k}^* \; v_{1,k,p,-\si}
+ \g_{1,-k,\si,\nu=4}^* \; \g_{1,p,\si,\nu=4}\; v_{1,-p} \; v_{1,-p-k,-\si}^*
 \big\}  \nonumber \\
+\; \g_{1,-k,\si,\nu=2}^* \; \g_{1,p,\si,\nu=2} \;
s_{2,k,-\si}^* \; s_{2,k,p,-\si}
+ \g_{1,-k,\si,\nu=3}^* \; \g_{1,p,\si,\nu=3} \;
s_{2,-p,-\si} \; s_{2,-p,-k,-\si}^*.
\;\;, \;\; 
\eea
\bea
\langle \cd_{2,-k,\si} \; c_{1,p,\si} \rangle|_{\si=\ua} =
- \; \text{sgn}(\si) \big\{    
\g_{2,-k,\si,\nu=1}^* \; \g_{1,p,\si,\nu=1} \; v_{2,-k}^* \; s_{2,-k,-p,-\si} 
+  \g_{2,-k,\si,\nu=3}^* \; \g_{1,p,\si,\nu=3} 
\; v_{1,p} \; s_{1,p,k,-\si}^*  \big\}\nonumber \\
+ \; \g_{2,-k,\si,\nu=2}^* \; \g_{1,p,\si,\nu=2} \; 
s_{1,-k,-\si}^* \; v_{1,-k,-p,-\si} 
+ \g_{2,-k,\si,\nu=4}^* \; \g_{1,p,\si,\nu=4} \; 
v_{2,p,k,-\si}^* \; s_{2,p,k,-\si}
\;\;, \;\;  
\eea
\bea
\langle \cd_{2,-k,\si} \; c_{1,p,\si} \rangle|_{\si=\da} =
- \; \text{sgn}(\si) \big\{    
\g_{2,-k,\si,\nu=2}^* \; \g_{1,p,\si,\nu=2} \; v_{2,k}^* \; s_{2,k,p,-\si} 
+  \g_{2,-k,\si,\nu=4}^* \; \g_{1,p,\si,\nu=4} 
\; v_{1,-p} \; s_{1,-p,-k,-\si}^*  \big\}  \nonumber \\
+ \; \g_{2,-k,\si,\nu=1}^* \; \g_{1,p,\si,\nu=1} \; 
s_{1,k,-\si}^* \; v_{1,k,p,-\si} 
+ \g_{2,-k,\si,\nu=3}^* \; \g_{1,p,\si,\nu=3} \; 
v_{2,-p,-k,-\si}^* \; s_{2,-p,-k,-\si}
\;\;, \;\;   \label{cdw2}
\eea
The asymmetry in these indices $\nu$ in the matrix elements follows the
asymmetry of $\nu$ in $M_{k,p}^{\dagger}$ above.

But we have $
\langle   \cd_{i,-k,\si} c_{i,p,-\si}   \rangle =
\langle   \cd_{i,-k,\si} c_{j,p,-\si}   \rangle = 0$.
However, upon introducing spin-triplet pairing terms such as
\be
 w_{i,k,p,\si} \; \cd_{i,k,\si,\nu} \cd_{i,p,\si,\nu}  
+ t_{i,k,p,\si} \; \cd_{i,k,\si,\nu} \cd_{j,p,\si,\nu}  \;\;, \label{eqtr}
\ee
etc. in $A_{i,k,p,\nu}^{\dagger}$ - also c.f. Appendices A and B -
we obtain non-zero matrix elements
\bea
\langle   \cd_{i,-k,\si} \; c_{i,p,-\si}   \rangle \propto  \{
s_{j,k,-\si} \; t_{j,k,p,-\si} , \; s_{j,p,\si}  \; t_{j,p,k,\si} , \; 
  v_{i,p} \; w_{i,p,k,\si}, \; v_{i,k} \; w_{i,k,p,-\si} \}  
  \;\;  , \;\;   \label{cdw3}  \\
\langle   \cd_{i,-k,\si} \; c_{j,p,-\si}   \rangle \propto \{ 
s_{j,k,-\si} \; w_{j,k,p,-\si} , \; s_{i,p,\si}  \; w_{i,p,k,\si} , \;
 v_{j,p} \; t_{j,p,k,\si}, \; v_{i,k} \; t_{i,k,p,-\si}  \}
   \label{cdw4} \;\;  . \;\;  
\eea

We do not provide a numerical evaluation of eqs. (\ref{cdw1})-(\ref{cdw2}),
(\ref{cdw3}), (\ref{cdw4}). The solution for $\Psi$ in eq. (\ref{eqpsi}) 
requires additional
algorithmic and programming effort, which is left for future work.

We note that {\em in principle} it is possible to have non-zero
expectation values for charge and spin density 
$\langle \cd_{i,k+Q,\si} c_{j,k,\pm \si}\rangle$ 
- with both $i=j$ and $i\neq j$ - for some $Q$-range \cite{pse},
while the anomalous propagators of the theory
$F_{i,\si}^{\dagger}, F_{ij,\si}^{\dagger}$ - c.f. 
eqs. (\ref{anf1}), (\ref{anf2}) - are {\em zero}. 
This regime may be relevant for  
the pseudogap phase of the copper oxide superconductors \cite{tim}.
Recent experimental works probing a CDW order in the pseudogap phase of the
cuprates include \cite{cdwe1,cdwe2,cdwe3}, and relevant theoretical proposals
include \cite{cdwt,cdwt2,cdwt3}.

In \cite{lut,psa} (treating different models though) the coexistence 
of charge and spin density wave order with superconductivity was explored.
We emphasize that, as far as we can currently see, this coexistence 
is {\em not compulsory}, though possible, in our approach.

\vspace{0.6cm}
\centerline{\bf 7. Summary }
\vspace{0.2cm}

In summary, using the new fermion quantum index, a variational fermionic
wavefunction $\Psi$, sustaining superfluidity, was introduced. 
Two different spin triplet versions of $\Psi$ can be found in Appendices
A and B.  
$\Psi$ accounts both for the 
intra-species $V_{i,q}$ and inter-species interactions $F_q$, with
an arbitrary momentum dependence, in an equally rigorous and comprehensive
manner. In the frame of this strong coupling approach, $\Psi$ can 
also yield finite charge and/or spin density wave order, irrespectively of
the existence of superconductivity in the system. The ground states,
for appropriate interspecies potential $F_q$, have an unusual Fermi
occupation factor equal to 1/2, deep in the Fermi sea. This is valid
{\em both} for the normal and the superfluid state
for the case of equivalent spin up and down fermions, and can be 
understood as a minimization of the kinetic energy effect. It should be
possible to check this prediction of the theory against experiments
which probe, in an {\em unbiased} manner, the fermion occupation in the 
core of the Fermi sea.

Also, at the end of Section 4 we point out that this unusual Fermi 
occupation factor
of 1/2 happens to satisfy relevant constraints for systems with a finite 
number of electrons.
Interestingly, these constraints were derived in \cite{db,kly}, 
in a context totally independent from ours.

\vspace{0.6cm}
\centerline{\bf Acknowledgments }
\vspace{0.2cm}

The author is indebted to Gregory Psaltakis and Ioannis Smyrnakis for 
invaluable discussions. Comments by Konstantinos Moulopoulos and Jiannis 
Pachos are acknowledged, as well as discussions on numerical methods with 
Georgios Zouraris. Peter Kopietz provided constructive criticism.

\vspace{6cm}
\centerline{\bf Appendix A: A spin triplet state with equal spin pairing}
\vspace{0.2cm}

We introduce a spin triplet version of the wavefunction $\Psi$, which 
corresponds to the "equal spin pairing" (ESP) case with parallel pair spins
only. In principle, the ESP state does not yield the lowest lying ground 
state \cite{bw} for the single species case, which may apply here as well.

To begin with, denoting by $\delta$ the new index, we introduce
\be
B_{i,k,\si,\de}^{\dagger} = u_{i,k,\si} +
w_{i,k,\si} \; \cd_{i,k,\si,\de} \; \cd_{i,-k,\si,\de}
+ t_{+,i,k,\si} \; \cd_{i,k,\si,\de} \; \cd_{j,-k,\si,\de}
+ t_{-,i,k,\si} \; \cd_{i,-k,\si,\de} \; \cd_{j,k,\si,\de}
\;\;. \label{eqbda}
\ee
$B_{i,k,\si,\de}^{\dagger}$ is a bosonic operator, creating spin triplet
pairs of fermions, and $(i,j)=\{(1,2),(2,1)\}$.

Henceforth we divide the momentum space into two parts, say $k>0$ (sgn$(k)=+$)
and $k<0$ (sgn$(k)=-$).
For $k>0$ we form the following multiplet of $B_{i,k,\si,\de}^{\dagger}$'s
\be
M_k^{\dagger}  = B_{1,k,\ua,\de=1}^{\dagger} \; B_{1,k,\da,\de=1}^{\dagger} \;
B_{2,k,\ua,\de=2}^{\dagger} \; B_{2,k,\da,\de=2}^{\dagger} \;\;. \label{mtr1}
\ee
This multiplet creates all states with momenta $\pm k$, and we take
$N_o=2$ as in section II.

Now we introduce the disentangled state
\be
|\Psi\rangle  = \prod_{k>0} M_k^{\dagger} \; |0\rangle  \;\;.
\ee
Note that {\em all} $B_{i,k,\si,\de}^{\dagger}$'s in $|\Psi\rangle$ 
{\em commute with each other.}

The normalization $\langle \Psi|\Psi\rangle =1$ implies
\be
|u_{i,k,\si}|^2+|w_{i,k,\si}|^2+|t_{+,i,k,\si}|^2+|t_{-,i,k,\si}|^2=1 \; . \;
\ee
Fermion statistics yields $w_{i,-k,\si}=-w_{i,k,\si}$.

Calculations are straightforward for the matrix elements derived
from $|\Psi\rangle$. For 2 fermion species with dispersions
$\ep_{i,k,\si}=\ep_{i,-k,\si}$ and for $k>0$ we have
\bea
 c_{1,k,\ua} \; M_k^{\dagger} \; |0\rangle = \g_{1 k \ua,1} \;
(w_{1,k,\si} \; \cd_{1,-k,\da,\de=1} +  t_{+,1,k,\ua} \; \cd_{2,-k,\da,\de=1})
\; B_{1,k,\da,\de=1}^{\dagger} \;
B_{2,k,\ua,\de=2}^{\dagger} \; B_{2,k,\da,\de=2}^{\dagger} 
\; |0\rangle \nonumber \\
- \g_{1 k \ua,2} \;
t_{-,2,k,\ua} \;  \cd_{2,-k,\da,\de=2} \; B_{1,k,\ua,\de=1}^{\dagger} \; 
B_{1,k,\da,\de=1}^{\dagger} \; B_{2,k,\da,\de=2}^{\dagger} \; |0\rangle \;\;.
\eea
Then
\be
\langle 0| \; M_k \; \cd_{1,k,\ua} c_{1,k,\ua} \; M_k^{\dagger} \; |0\rangle =
\g_{1 k \ua,1}^2 \; ( |w_{1,k,\ua}|^2+|t_{+,1,k,\ua}|^2 )
+|\g_{1 k \ua,2}|^2 \; |t_{-,2,k,\ua}|^2  \;\;. \;\;
\ee
Likewise,
\be
\langle 0| \; M_k \; \cd_{2,k,\ua} c_{1,k,\ua} \; M_k^{\dagger} \; |0\rangle =
-  \g_{2 k \ua,1}^* \; \g_{1 k \ua,1} \; t_{-,1,k,\ua}^* \;w_{1,k,\ua}
- \g_{2 k \ua,2}^* \; \g_{1 k \ua,2} \; t_{-,2,k,\ua}\;w_{2,k,\ua}^* 
\;\;.\;\;
\ee
and
\be
\langle 0| \; M_k \; c_{2,-k,\da} c_{1,k,\ua} \; M_k^{\dagger} \; |0\rangle =
  \g_{2 -k \ua,1} \; \g_{1 k \ua,1} \; u_{1,k,\si}^* \;  t_{+,1,k,\si}
- \g_{2 -k \ua,2} \; \g_{1 k \ua,2} \; u_{2,k,\si}^* \;  t_{-,2,k,\si}  
 \;\;.\;\;
\ee

Using the commutativity of $B_{i,k,\si,\de}^{\dagger}$'s and generalizing
the previous equations, we obtain 
($\langle C \rangle =\langle \Psi | C | \Psi \rangle)$
\bea
n_{i,k,\si}  = \langle \cd_{i,k,\si} c_{i,k,\si} \rangle
= \g_{i k \si,i}^2 \; (|w_{i,k,\si}|^2+|t_{l_k,i,k,\si}|^2) 
+|\g_{i k \si,j}|^2 \; |t_{-l_k,j,k,\si}|^2
\;\;, \;\; (i,j)=(1,2),(2,1) \;\; , \\
\ze_{k,\si} = \langle \cd_{i,k,\si} c_{j,k,\si} \rangle
=-l_k \;(\g_{j k \si,i}^*\; \g_{i k \si,i} \; w_{i,k,\si}^* \;t_{-l_k,i,k,\si} 
+ \g_{j k \si,j}^*\; \g_{i k \si,j} \; w_{j,k,\si}^* \; t_{-l_k,j,k,\si} )  
\; \; ,  \\
\Gamma_{k,\si} = \langle c_{j,-k,\si} c_{i,k,\si} \rangle
=  \g_{j -k \si,i}^*\; \g_{i k \si,i} \; u_{i,k,\si}^* \;  t_{-l_k,i,k,\si}
- \g_{j -k \si,j}^* \; \g_{i k \si,j} \;u_{j,k,\si}^* \;  t_{l_k,j,k,\si} 
\;\; , \\
\Phi_{i,k,\si} = \langle c_{i,-k,\si} c_{i,k,\si} \rangle
= l_k \; \g_{i -k \si,i} \; \g_{i k \si,i} \; u_{i,k,\si}^* \; w_{i,k,\si}
 \;\; , \;\;
\eea
with $l_k = $sgn$(k)$.

A general Hamiltonian for two fermion species is given in eq. (\ref{ham1})
in section III.
Considering $\Psi$ above, we have 
for $\langle H\rangle =\langle \Psi|H|\Psi\rangle $, 
\bea
\langle H\rangle  = \sum_{i,k,\si} \xi_{i,k,\si} \; n_{i,k,\si} 
+ \sum_{k,\si}  \; h_k \; \big( \ze_{k,\si} +  \ze_{k,\si}^* \big) 
+ \frac{1}{2} \sum_{i,k,p,\si} \big( V_{i,q=0}-V_{i,k-p} \big)\; 
n_{i,k,\si} \;n_{i,p,\si}
\label{hams1} \\
+ \frac{1}{2} \sum_{i,k,p,\si} V_{i,k-p}\; \Phi_{i,k,\si}  \; \Phi_{i,p,\si}^*
-\sum_{k,p,\si} F_{k-p} \; \ze_{k,\si} \; \ze_{p,\si}^* 
+  F_{q=0} \; n_{1} \;n_{2}
+ \sum_{k,p,\si} F_{k-p} \; \Gamma_{k,\si} \; \Gamma_{p,\si}^*
\nonumber \;\;,  
\eea
with $(i,j)=\{(1,2),(2,1)\}$. The first term in the second line is exactly
the usual BCS-like pairing term, and the last term is 
the equivalent inter-species pairing term due to $F_q$.
We note the {\em formal equivalence} between $\langle H\rangle$ above
and $\langle H\rangle$ in eq. (\ref{ham2}) for the spin singlet case.

The minimization procedure for $\langle H\rangle$ and the finite temperature
extension proceed as shown in the main part of the paper for the spin 
singlet case.

\vspace{0.6cm}
\centerline{\bf Appendix B: A generic spin triplet state }
\vspace{0.2cm}

Herein we introduce a spin triplet version of the wavefunction $\Psi$, which 
is a generalization of the Balian-Werthamer state \cite{bw}, including
all three components of the total spin. In Appendix A we
introduced the ESP case with parallel pair spins
only.

First, denoting by $\delta$ the new index, we introduce
\bea
C_{i,k,\de}^{\dagger} = u_{i,k} 
+ v_{i,k} \;  \big( \cd_{i,k,\ua,\de} \; \cd_{i,-k,\da,\de}
 + \cd_{i,k,\da,\de} \; \cd_{i,-k,\ua,\de}  \big)
+ w_{i,k,\ua} \; \cd_{i,k,\ua,\de} \; \cd_{i,-k,\ua,\de}
+ w_{i,k,\da} \; \cd_{i,k,\da,\de} \; \cd_{i,-k,\da,\de} \label{eqcda} \\
+  s_{i,k} \; \big( \cd_{i,k,\ua,\de} \; \cd_{j,-k,\da,\de}
+ \cd_{i,k,\da,\de} \; \cd_{j,-k,\ua,\de} 
+ \cd_{i,-k,\ua,\de} \; \cd_{j,k,\da,\de}
+ \cd_{i,-k,\da,\de} \; \cd_{j,k,\ua,\de} \big)    \nonumber \\
+ t_{i,k,\ua} \;  \big( \cd_{i,k,\ua,\de} \; \cd_{j,-k,\ua,\de}
+ \cd_{i,-k,\ua,\de} \; \cd_{j,k,\ua,\de}  \big)
+ t_{i,k,\da} \;  \big( \cd_{i,k,\da,\de} \; \cd_{j,-k,\da,\de}
+ \cd_{i,-k,\da,\de} \; \cd_{j,k,\da,\de}  \big)
\;\;.   \nonumber
\eea
$C_{i,k,\de}^{\dagger}$ is a bosonic operator, creating spin triplet
pairs of fermions,
and $(i,j)=\{(1,2),(2,1)\}$. Other variants of $C_{i,k,\de}^{\dagger}$
can be envisaged as well.

Henceforth we divide the momentum space into two parts, say $k>0$ (sgn$(k)=+$)
and $k<0$ (sgn$(k)=-$).
For $k>0$ we form the following multiplet of $C_{i,k,\de}^{\dagger}$'s
\be
M_k^{\dagger}= C_{1,k,\de=1}^{\dagger} \; C_{2,k,\de=2}^{\dagger} \;\;.
 \label{mtr2}
\ee
This multiplet creates all states with momenta $\pm k$, and we take
$N_o=2$, as for the two other versions of $\Psi$ above.

Now we introduce the disentangled state
\be
|\Psi\rangle  = \prod_{k>0} M_k^{\dagger} \; |0\rangle  \;\;.
\ee
Note that {\em all} $C_{i,k,\de}^{\dagger}$'s in $|\Psi\rangle$ {\em commute 
with each other.}

The normalization $\langle \Psi|\Psi\rangle =1$ implies
\be
|u_{i,k}|^2+2|v_{i,k}|^2+|w_{i,k,\ua}|^2+|w_{i,k,\da}|^2+4|s_{i,k}|^2+
2|t_{i,k,\ua}|^2+2|t_{i,k,\da}|^2=1 \; . \;
\ee
Fermion statistics yields $v_{i,-k}=-v_{i,k}$ and $w_{i,-k,\si}=-w_{i,k,\si}$.

Calculations are straightforward for the matrix elements derived
from $|\Psi\rangle$. For 2 fermion species with dispersions
$\ep_{i,k,\si}=\ep_{i,-k,\si}$ and for $k>0$ we have
\bea
c_{1,k,\ua} \; M_k^{\dagger} \; |0\rangle = \g_{1 k \ua,1}
(v_{1,k} \; \cd_{1,-k,\da,\de=1} + w_{1,k,\ua} \; \cd_{1,-k,\ua,\de=1} 
+  s_{1,k} \; \cd_{2,-k,\da,\de=1} +  t_{1,k,\ua} \; \cd_{2,-k,\ua,\de=1})
\;  C_{2,k,\de=2}^{\dagger} 
\; |0\rangle \nonumber \\
- \g_{1 k \ua,2} \; (s_{2,k} \;  \cd_{2,-k,\da,\de=2} 
+t_{2,k,\ua} \;  \cd_{2,-k,\da,\de=2} ) \; C_{1,k,\de=1}^{\dagger} \;  
|0\rangle \;\;.
\eea
Then
\be
\langle 0| \; M_k \; \cd_{1,k,\ua} c_{1,k,\ua} \; M_k^{\dagger} \; |0\rangle =
|\g_{1 k \ua,1}|^2 ( |v_{1,k}|^2 + |w_{1,k,\ua}|^2 + |s_{1,k}|^2
+|t_{1,k,\ua}|^2)
+ |\g_{1 k \ua,2}|^2 (|s_{2,k}|^2  +|t_{2,k,\ua}|^2) \;\;. \;\;
\ee
Likewise,
\be
\langle 0| \; M_k \; \cd_{2,k,\ua} c_{1,k,\ua} \; M_k^{\dagger} \; |0\rangle =
-\g_{1 k \ua,1}\; \g_{2 k \ua,1}^* \;
(t_{1,k,\ua}^* \;w_{1,k,\ua} + v_{1,k}\; s_{1,k}^*) 
-\g_{1 k \ua,2}\; \g_{2 k \ua,2}^* \;
(t_{2,k,\ua}\;w_{2,k,\ua}^* + v_{2,k}^* \;s_{2,k} )  \;\;.\;\;
\ee
and
\be
\langle 0| \; M_k \; c_{2,-k,\ua} c_{1,k,\ua} \; M_k^{\dagger} \; |0\rangle =
\g_{1 k \ua,1}\; \g_{2 -k \ua,1} \; u_{1,k}^* \;  t_{1,k,\ua}
-  \g_{1 k \ua,2}\; \g_{2 -k \ua,2} \; u_{2,k}^* \;  t_{2,k,\ua} 
 \;\;.\;\;
\ee

Using the commutativity of $C_{i,k,\de}^{\dagger}$'s and generalizing
the previous equations, we obtain 
($\langle B \rangle =\langle \Psi | B | \Psi \rangle)$
\bea
n_{i,k,\si}  = \langle \cd_{i,k,\si} \; c_{i,k,\si} \rangle
= |\g_{i k \si,i}|^2 \;
( |v_{i,k}|^2 + |w_{i,k,\si}|^2+|s_{i,k}|^2 +|t_{i,k,\si}|^2)
+|\g_{i k \si,j}|^2 \; (|s_{j,k}|^2 + |t_{j,k,\si}|^2)
\;\;, \\
\ze_{k,\si} = \langle \cd_{i,k,\si}  \; c_{j,k,\si} \rangle
= - l_k \; \{\g_{i k \si,i}^* \; \g_{j k \si,i} \;
(w_{i,k,\si}^* \; t_{i,k,\si} +v_{i,k}^* s_{i,k}) 
+\g_{i k \si,j}^* \; \g_{j k \si,j} \;
(w_{j,k,\si} \; t_{j,k,\si}^* + v_{j,k} s_{j,k}^* )\}  \; \; ,  \\
\la_{k,\si} = \langle c_{2,-k,\si} c \; _{1,k,\si} \rangle
= \g_{1 k \si,1} \; \g_{2 -k \si,1} \; u_{1,k}^* \;  t_{1,k,\si}
- \g_{1 k \si,2} \; \g_{2 -k \si,2} \; u_{2,k}^* \;  t_{2,k,\si} \; \; ,\\
g_{k,\si} = \langle c_{2,-k,-\si}  \; c_{1,k,\si} \rangle
=  \g_{1 k \si,1} \; \g_{2 -k -\si,1} \;u_{1,k}^* \;s_{1,k} 
-\g_{1 k \si,2} \; \g_{2 -k -\si,2} \; u_{2,k}^* \;s_{2,k}   \;\; , \\
b_{i,k,\si} = \langle c_{i,-k,-\si}  \; c_{i,k,\si} \rangle
= l_k \; \g_{i k \si,i} \; \g_{i -k -\si,i} \;u_{i,k}^* \; v_{i,k} \;\; ,\;\;
 d_{i,k,\si} = \langle c_{i,-k,\si}  \; c_{i,k,\si} \rangle
= l_k \; g_{i k \si,i} \; \g_{i -k \si,i}  \; u_{i,k}^* \; w_{i,k,\si} 
 \;\; , \;\;
\eea
with $l_k = $sgn$(k)$ and $(i,j)=(1,2),(2,1)$.

A general Hamiltonian for two fermion species is given in eq. (\ref{ham1})
in section III.
Considering $\Psi$ above, we have 
for $\langle H\rangle =\langle \Psi|H|\Psi\rangle $,
\bea
\langle H\rangle  = \sum_{i,k,\si} \xi_{i,k,\si} \; n_{i,k,\si} 
+ \sum_{k,\si}  \; h_k \; \big( \ze_{k,\si} +  \ze_{k,\si}^* \big) 
+ \frac{1}{2} \sum_{i,k,p,\si} \big( V_{i,q=0}-V_{i,k-p} \big)\; 
n_{i,k,\si} \;n_{i,p,\si}
+  F_{q=0} \; n_{1} \; n_{2}
 \nonumber \\
+ \frac{1}{2} \sum_{i,k,p,\si} V_{i,k-p}\; \big( b_{i,k,\si} b_{i,p,\si}^* \; 
+ d_{i,k,\si}^* d_{i,p,\si}   \big)
-\sum_{k,p,\si} F_{k-p} \; \ze_{k,\si} \; \ze_{p,\si}^* 
+ \sum_{k,p,\si} F_{k-p} \; \big( \la_{k,\si} \; \la_{p,\si}^*
+ g_{k,\si} \; g_{p,\si}^* \big) \;\;, \;\;  \label{hams2}  
\eea
with $(i,j)=\{(1,2),(2,1)\}$. The first term in the second line is exactly
the usual BCS-like pairing term, and the last term is 
the equivalent inter-species pairing term due to $F_q$.
Allowing for pairs with non-zero total momentum in $|\Psi\rangle $,
as shown in Section VI, yields additional terms in $\langle H\rangle $.
In general, and as already noted, expanding the Hilbert space of 
$|\Psi\rangle $ via the 
inclusion of more pairing correlations than the ones shown, may
lead to a {\em further reduction of the ground state energy}.

The minimization procedure for $\langle H\rangle$ and the finite temperature
extension proceed as shown in the main part of the paper for the spin 
singlet case.

\vspace{.6cm}
\centerline{\bf Appendix C: Energy minimization conditions}
\vspace{.3cm}

Here we give the explicit expressions, from which the variables
$ \theta_{i,k},\phi_{i,k}, a_{i,k}, b_{i,k}$ and $\om_{i,k}$ are calculated.
They are the explicit forms of eqs. (\ref{eqm1}), which correspond to
a minimum of the total energy $E$ with respect to these variables.

Some relevant parameter definitions were given in eq. (\ref{eqom}).

The condition $\partial E/\partial \theta_{i,k} = 0$
yields
\bea
0=\g_{i k, i} \; \cos(\phi_{i,k}) \; 
\Big [  \g_{i k, i} \;\cos(\phi_{i,k}) \; \Xi_{i,k} \; \sin(2 \th_{i,k})
+ \sqrt{2} \; \sin(\phi_{i,k}) \; \sin(\eta_{j,k})\; \text{Re} \{
 \exp(i \; (-1)^i \; \Omega_{ij,k}) \; [h_k - S_k^* ] \} \; \cos(\th_{i,k})
  \nonumber   \\
- \sqrt{2} \; \sin(\phi_{i,k}) \; 
\text{Re} \{ \g_{j k, i} \; \exp(i b_{i,k}) \;
 \Phi_k^* \} \; \sin(\th_{i,k})  - \g_{i k, i} \;\cos(\phi_{i,k}) \; 
\text{Re} \{ \Delta_{i,k}\; \exp(-i a_{i,k})\} \; \cos(2 \th_{i,k}) \Big ] 
\;\; . \;\;
\eea

The minimization condition $\partial E /\partial \phi_{i,k}=0$
yields
\be
\phi_{i,k} = -\frac{1}{2} \arctan \left( \frac{C_{i,k}}{D_{i,k}} \right) \;\;, 
\;\;
\ee
with 
\bea
C_{i,k} =\sqrt{2} \; \g_{i k, i} \; \Big [
\cos(\theta_{i,k}) \; \text{Re} \{ \g_{j k, i}^* \; \exp(i b_{i,k}) \;
 \Phi_k^* \} 
+ \sin(\theta_{i,k}) \;  \sin(\eta_{j,k}) \; \text{Re} \{  
 \exp(i \; (-1)^i \; \Omega_{ij,k})  \; ( h_k - S_k^*) \} 
\Big ]
\eea
and
\be
D_{i,k} = \g_{i k, i}^2 \; \Xi_{i,k} \; \{  1/2 -\sin(\theta_{i,k})^2 \}
+ |\g_{j k, i}|^2 \; \Xi_{j,k} /2  
+ \g_{i k, i}^2 \; \sin(2 \theta_{i,k}) \;
\text{Re} \{ \Delta_{i,k}\; \exp(-i a_{i,k}) \} /2 \;.
\ee
The correction $\de \phi_{i,k}$ - c.f. Sections IV and V - is given by
\be
\de \phi_{i,k} = C_{i,k}/(2 D_{i,k}) \;\;. \label{def}
\ee
Likewise, the correction $\de \eta_{i,k}$ is given by
\be
\de \eta_{i,k} = 2 \text{Re} \{T_{i,k}\}/K_{i,k} \;\;,  \label{dee}
\ee
with $K_{i,k}= 2 \Xi_{i,k} \; [ |s_{j,k}|^2  - (|v_{i,k}|^2 + |s_{i,k}|^2)]
+ 2 \text{Re} ( \Delta^{*}_{i,k} \; u_{i,k} v_{i,k})$.

We note that, for $|\phi_{i,k}|,|\eta_{i,k}| \rightarrow \pi/2$, 
both $\de \phi_{i,k},\de \eta_{i,k}$ are {\em very small}.

Minimization with regard to $a_{i,k},b_{i,k},\om_{i,k}$ yields similar
equations. $\partial E /\partial a_{i,k}=0$
yields
\be
a_{i,k} = - \arctan \left( \frac{ \text{Im} 
\{ q_{i,k} \; \Delta^{*}_{i,k} + (-1)^i \; r_{i,k} \} }
{\text{Re}
\{ q_{i,k} \; \Delta^{*}_{i,k} - r_{i,k} \} }
\right) \;\;, 
\;\;
\ee
with 
\bea
q_{i,k} = \sin(2 \theta_{i,k})  \cos^2(\phi_{i,k})  \cos^2(\eta_{i,k})/2
\;\; , \;\;
r_{i,k} = x_{0,i,k} \; S^*_{k} \; 
\exp(i \; (-1)^i \; [b_{i,k}+\om_{j,k}]) \;\; , \;\;
\nonumber  \\
x_{0,i,k} = \sin(\theta_{i,k}) \;
\sin(2 \phi_{i,k}) \cos(\eta_{i,k}) \sin(\eta_{j,k}) /(2 \sqrt{2})  \;\;. 
\eea

 $\partial E /\partial b_{i,k}=0$
yields
\be
b_{i,k} = - \arctan \left( \frac{ \text{Im} 
\{ w_{i,k} + (-1)^i \; y_{i,k} \} }
{\text{Re}
\{ w_{i,k} + y_{i,k} \} }
\right) \;\;, \;\;
\ee
with 
\be
y_{i,k} = (-1)^i \; (h_k - S^*_{k}) \; x_{0,i,k} \; 
\exp(i \; (-1)^i \; [\om_{j,k} - a_{i,k}] ) \;\;, \;\;
w_{i,k} = \Phi^*_{k} \; \cos(\theta_{i,k}) \;
x_{0,i,k} \; \exp{(i \om_{j,k})} \; \; .\;\;
\ee

 $\partial E /\partial \om_{i,k}=0$
yields
\be
\om_{i,k} = \arctan \left( \frac{ \text{Im} 
\{ A_{i,k} - B_{i,k} \} }
{\text{Re}
\{ B_{i,k} + (-1)^i A_{i,k} \} }
\right) \;\;, \;\;
\ee
with 
\bea
A_{i,k} = (-1)^i \;  (h_k - S^*_{k}) \; x_{0,j,k} \; 
\exp(i \; (-1)^j \; [b_{j,k} - a_{j,k}] ) \;\;, \;\;  \\  \nonumber
B_{i,k}= \Phi^*_{k} \; \cos(\theta_{j,k}) 
\exp(i b_{j,k}) \; 
\sin(2 \phi_{j,k}) \cos(\eta_{j,k}) \sin(\eta_{i,k}) /(2 \sqrt{2}) \; \;\; .
\eea
In $A_{i,k}$ and $B_{i,k}$ most of the indices are indeed "$j$".

\vspace{0.6cm}
\centerline{\bf Appendix D: Details of the finite temperature 
dependence of the theory}
\vspace{0.3cm}

The finite temperature dependence of the theory can be derived 
through the equations of motion 
formalism for the Green's functions \cite{agd,psa} 
($\partial_{\tau} c_x(\tau) = [H,c_x(\tau)]$). We consider
the Green's functions
\bea
G_{i,\si}(k,\dt)&=& - \langle T_o \; c_{i,k,\si}(\tau) \cd_{i,k,\si}(\tau')
\rangle  \;\;, \\
G_{ij,\si}(k,\dt)&=& -\langle T_o \; c_{i,k,\si}(\tau) \cd_{j,k,\si}(\tau')
\rangle  \;\;, \\
F_{i,\si}^{\dagger} (k,\dt) &=&     \label{anf1}
\langle T_o \; \cd_{i,k,\si}(\tau) \cd_{i,-k,-\si}(\tau')\rangle  \;\;, \\
F_{ij,\si}^{\dagger} (k,\dt) &=&     \label{anf2}
\langle T_o \; \cd_{j,k,\si}(\tau) \cd_{i,-k,-\si}(\tau')\rangle  \;\;,
\eea
with $(i,j)=\{(1,2),(2,1)\}$ and $T_o$ denoting imaginary time ordering.

We obtain the exact coupled equations
\bea
\delta(\dt) = -(\partial_{\tau} +\xi_{i,p,\si}) G_{i,\si}(p,\dt)  
+ \sum_{k,q,\si'} V_{i,q} \langle T_o  \; \cd_{i,k-q,\si'}(\tau)  
c_{i,k,\si'}(\tau) c_{i,p-q,\si}(\tau)  \cd_{i,p,\si}(\tau') \rangle   
\nonumber  \\
- h_p  G_{ji,\si}(p,\dt) 
+ \sum_{k,q,\si'} F_{q} \langle T_o  \;  \cd_{j,k-q,\si'}(\tau)  
c_{j,k,\si'}(\tau) c_{i,p-q,\si}(\tau)  \cd_{i,p,\si}(\tau') \rangle \;\;,  \\  
0= (-\partial_{\tau} +\xi_{i,p,\si}) F_{i,\si}^{\dagger}(p,\dt) 
+ \sum_{k,q,\si'} V_{i,q} \langle T_o  \;  \cd_{i,p-q,\si}(\tau)  
\cd_{i,k+q,\si'}(\tau) c_{i,k,\si'}(\tau)  \cd_{i,-p,-\si}(\tau') \rangle   
\nonumber \\  
+ h_p F_{ij,\si}^{\dagger}(p,\dt)  
+ \sum_{k,q,\si'} F_{q} \langle T_o  \;  \cd_{i,p-q,\si}(\tau)  
\cd_{j,k+q,\si'}(\tau) c_{j,k,\si'}(\tau)  \cd_{i,-p,-\si}(\tau') \rangle \;\;,
 \\  
0 = - (\partial_{\tau} +\xi_{j,p,\si}) G_{ji,\si}(p,\dt)  
+ \sum_{k,q,\si'} V_{j,q} \langle T_o  \; \cd_{j,k-q,\si'}(\tau)  
c_{j,k,\si'}(\tau) c_{j,p-q,\si}(\tau)  \cd_{i,p,\si}(\tau') \rangle  
 \nonumber  \\
- h_p  G_{i,\si}(p,\dt) 
+ \sum_{k,q,\si'} F_{q} \langle T_o  \;  \cd_{i,k-q,\si'}(\tau)  
c_{i,k,\si'}(\tau) c_{j,p-q,\si}(\tau)  \cd_{i,p,\si}(\tau') \rangle   
\;\;,  \\  
0= (-\partial_{\tau} +\xi_{j,p,\si}) F_{ij,\si}^{\dagger}(p,\dt) 
+ \sum_{k,q,\si'} V_{j,q} \langle T_o  \;  \cd_{j,p-q,\si}(\tau)  
\cd_{j,k+q,\si'}(\tau) c_{j,k,\si'}(\tau)  \cd_{i,-p,-\si}(\tau') \rangle   
\nonumber  \\  
+ h_p F_{i,\si}^{\dagger}(p,\dt) 
+ \sum_{k,q,\si'} F_{q} \langle T_o  \;  \cd_{j,p-q,\si}(\tau)  
\cd_{i,k+q,\si'}(\tau) c_{i,k,\si'}(\tau)  \cd_{i,-p,-\si}(\tau') \rangle   
\;\;.  
\eea

We Fourier transform these equations from $\tau$ to the Matsubara 
energy $\ep_n=(2n+1)\pi T$, $T$ being the temperature, and we solve them
within the Hartree-Fock-Bogoliubov approximation.

We have the relevant factors, for which 
we {\em suppress} the labels $k,\si$
  
\bea
A_1= \sum_{p,q,\si'} V_{1,q} \big\{ -\de_{q,0} \; n_{1,p,\si'}
+ \de_{\si,\si'} \de_{k,p} \; n_{1,k-q,\si} \big\}
-F_{q=0} \; n_{2} \;\; , \;\; 
B_1 = \Delta_{1,k} \;\; , \;\; C_0= \sum_{q} F_q \; z_{k+q,\si} \;\;, \;\;
 \nonumber  \\
C_1= -h_k - C_0 \;\;,  \;\;
 D_1= \sum_{q} F_q \; g_{k+q,\si}  \;\;,  \;\;
 D_2= h_k + C_0^* \;\;,  \;\; A_3 = - h_k + C_0^* \;\;, \;\;
\nonumber  \\
 C_3=\sum_{p,q,\si'} V_{2,q} \big\{ -\de_{q,0}  \; n_{2,p,\si'} 
 +\de_{\si,\si'} \de_{k,p}  \; n_{2,k-q,\si}  \big\}
-F_{q=0} \; n_{1} \;\; , \;\; 
 D_3= -\Delta_{2,k} \;\;, \;\;
B_4 = h_k - C_0 \;\;. 
\eea

The set of equations for the normal and anomalous Green's functions,
depending on $k$ and $\ep_n$, is
\bea
1=(i \ep_n-\xi_{1,k,\si}) \; G_{1,\si}(k,\ep_n) + A_1 \; G_{1,\si}(k,\ep_n) 
+ B_1\; F_{1,\si}^{\dagger}(k,\ep_n)
+C_1 \; G_{21,\si}(k,\ep_n) + D_1 \; F_{12,\si}^{\dagger}(k,\ep_n)  
\;\;,  \\
0=(i \ep_n+\xi_{1,k,\si}) \;F_{1,\si}^{\dagger}(k,\ep_n) 
+ B_1^* \; G_{1,-\si}(k,\ep_n)  
 -A_1 \; F_{1,\si}^{\dagger}(k,\ep_n)
+D_1^* \; G_{21,-\si}(k,\ep_n) + D_2 \; F_{12,\si}^{\dagger}(k,\ep_n)  
\;\;,  \\
0=(i \ep_n - \xi_{2,k,\si}) \;  G_{21,\si}(k,\ep_n) 
+ A_3 \; G_{1,\si}(k,\ep_n) + 
D_1 \; F_{1,-\si}^{\dagger}(k,\ep_n)
+C_3 \; G_{21,\si}(k,\ep_n) + D_3 \; F_{12,-\si}^{\dagger}(k,\ep_n)  
\;\;,  \\
0=(i \ep_n+\xi_{2,k,\si}) \;F_{12,\si}^{\dagger}(k,\ep_n) 
+ D_1^* \; G_{1,\si}(k,\ep_n) 
+ B_4  \; F_{1,\si}^{\dagger}(k,\ep_n)
+D_3^* \; G_{21,-\si}(k,\ep_n) - C_3 \; F_{12,\si}^{\dagger}(k,\ep_n)   
\;\;,
\eea
and also the equivalent set with the indices 1 and 2 interchanged.
For the case of equivalent up and down spin 
$\xi_{i,k,-\si}=\xi_{i,k,\si}=\xi_{i,k}$ for both fermion species, 
the solutions are

\vspace{.3cm}

\be 
G_{1}(k,i\ep_n)=\frac{Z_1(k,i\ep_n)}{D(k,i \ep_n)}, \;\;
F_{1}^{\dagger}(k,i\ep_n)=\frac{X_1(k,i\ep_n)}{D(k,i \ep_n)},  \;\;
G_{21}(k,i\ep_n)=\frac{Z_{21}(k,i\ep_n)}{D(k,i \ep_n)},  \;\;
F_{12}^{\dagger}(k,i\ep_n)=\frac{X_{12}(k,i\ep_n)}{D(k,i \ep_n)}\;,
\label{feq}
\ee
and likewise for $G_{2}(k,i\ep_n),F_{2}^{\dagger}(k,i\ep_n),G_{12}(k,i\ep_n)$
and $ F_{21}^{\dagger}(k,i\ep_n) $. Setting
\be
R_1 = (\xi_{1,k}-A_1)^2+|B_1|^2 \;\;,\;\; R_2 = (\xi_{2,k}-C_3)^2+|D_3|^2
\;\; ,
\ee
the numerators 
$Z_1(k,i\ep_n)$, $X_1(k,i\ep_n)$, $Z_{21}(k,i\ep_n)$, $X_{12}(k,i\ep_n)$ are

\bea
Z_1 =-i \ep_n^3+\ep_n^2 \; (A_1-\xi_{1,k})-i \ep_n \;
(R_2 + D_2 B_4 + |D_1|^2)
 +D_1 D_2 D_3^* + D_1^* D_3 B_4   \nonumber  \\  
+ (A_1-\xi_{1,k}) R_2 
+ (C_3-\xi_{2,k}) (|D_1|^2 - D_2 B_4)
  \;, \;\; 
\eea

\bea
X_1 = B_1^* \; \ep_n^2 
+ i \ep_n \; D_1^* (A_3+D_2)  + B_1^* R_2 + \xi_{2,k} \; D_1^* (A_3-D_2)
-A_3 (C_3 D_1^*+D_2 D_3^*) + D_1^* D_2 C_3 - (D_1^*)^2 D_3
 \;,  \;\; \label{areq}
\eea

\bea
Z_{21} = A_3 \; \ep_n^2
 + i \ep_n \; [B_1^* D_1 + D_1^* D_3 + A_3 (A_1-\xi_{1,k}+C_3-\xi_{2,k})]
+ (A_1+\xi_{1,k}) [ D_1^* D_3 + A_3 (C_3-\xi_{2,k}) ]   \nonumber  \\  
+ \xi_{2,k} \; B_1^* D_1  - |D_1|^2 D_2 + D_2 A_3 B_4 - B_1^*(D_3 B_4+D_1 C_3)
 \;, \;\;  
\eea

\bea
X_{12} = D_1^* \ep_n^2 
+i \ep_n \;[ D_1^* (A_1-\xi_{1,k}-C_3+\xi_{2,k}) +B_1^* B_4+A_3 D_3^*]
+D_1^* (|D_1|^2 -A_3 B_4)
\nonumber   \\ 
+ (A_1-\xi_{1,k}) [D_1^* (C_3-\xi_{2,k}) -A_3 D_3^*]
+ B_1^* [ B_4 (C_3-\xi_{2,k}) - D_1 D_3^* ]
\;. \; \;
\eea

The denominator is (now with $ i \ep_n \rightarrow \ep$)
\be
D(k,\ep)=\ep^4+Q \; \ep^2 +S \; \ep + Y \;\;,
\ee
with
\bea
Q=-R_1-R_2-2|D_1|^2-D_2 B_4  \;\;, \; \;
S = (A_1 -\xi_{1,k} + C_3-\xi_{2,k}) (C_1 A_3 -D_2 B_4)  \;\;,\;\;
\nonumber  \\
Y=R_1 \; R_2 
+(A_1-\xi_{1,k}) (C_3-\xi_{2,k}) \; (2 |D_1|^2 -C_1 A_3  -D_2 B_4)
  \\
+(A_1-\xi_{1,k}) \; [B_1^* D_1 (B_4-C_1) -D_1 D_2 D_3^*]
+(C_3-\xi_{2,k}) \; D_1^* B_1 \; (D_2+A_3)
\nonumber  \\
+A_3 D_3^* \; [D_1 \xi_{1,k} -D_2(B_1+D_1)]
+(A_1-\xi_{1,k}) \; D_1^* D_3  (B_4-C_1) -B_1^* D_1^2 D_3^* 
\nonumber  \\
-B_1 (D_1^*)^2 D_3  - |D_1|^2 \; (A_3 B_4 - C_1 D_2)- D_2 A_3 D_3^*\;(B_1+D_1)
+C_1 B_4 \; (D_2 A_3 - B_1^* D_3) 
\;\; \nonumber .
\eea

Then $D(k,E_k)=0$ yields four solutions for the quasiparticle energies
$E_k$. Setting 
$K=27 S^2-72 Y Q+2Q^3, L=12 Y+Q^2, N=K-\sqrt{K^2-4 L^3}$ and 
$W=\{ -2 Q+L (2/N)^{1/3}+(N/2)^{1/3} \}/3$ we have
\bea
E_{(a,b),k} = \frac{1}{2}\sqrt{W} \pm  
\frac{1}{2}\sqrt{ -W-2Q-2 S/\sqrt{W}  }  \;\;, \;\;
E_{(c,d),k} = -\frac{1}{2}\sqrt{W} \pm  
\frac{1}{2}\sqrt{ -W-2Q+2 S/\sqrt{W}  }  \;\;.  \label{eqdis}
\eea

They depend implicitly on the temperature $T$ through the factors
$u_{i,k}(T),v_{i,k}(T),s_{i,k,\sigma}(T)$. The latter can be calculated
by noting that 
\bea
G_{i,\si}(k,\tau=0) = - \langle c_{i,k,\si} \cd_{i,k,\si} \rangle 
= T \; \sum_{\ep_n} G_{i,\si}(k,\ep_n) \;\;,  \\
F_{i,\si}^{\dagger}(k,\tau=0) = \langle \cd_{i,k,\si} \cd_{i,-k,-\si}\rangle 
= T \; \sum_{\ep_n} F_{i,\si}^{\dagger}(k,\ep_n) \;\;, \label{eqftc}
\eea
and through the use of eqs. (\ref{feq}). For $T\leq T_c$, the 
critical temperature, 
$F_{i,\si}^{\dagger}(k)$ - and possibly $F_{ij,\si}^{\dagger}(k)$ - becomes 
non-zero.
{\em This is how $T_c$ can be calculated in this theory.}

\vspace{.3cm}

In the {\em numerical solutions} for the new ground states thus far obtained, 
the angles $\eta_{i,k}$ and $\phi_{i,k}$
take exclusively the values $|\eta_{i,k}|=0, \pi/2-\de \eta_{i,k}$ and
$|\phi_{i,k}|=0, \pi/2-\de \phi_{i,k}$, with both 
$|\de \eta_{i,k}|, \; |\de \phi_{i,k}|$ being {\em very small}.
Taking also the hybridization $h_k$=0, we have 
$C_1,D_1,D_2,A_3,B_4 \rightarrow 0$. However, the gaps $\De_{i,k}$ are
perfectly finite.
Then, 
\be
Q=-R_1-R_2 \;\; , \;\; S=0  \;\; , \;\; Y=R_1 \; R_2  \;\; . \;\;
\ee
Hence we obtain two (double) branches for the quasiparticle dispersion
\be
E_{1,k}^2=R_1 \;\; , \;\;  E_{2,k}^2=R_2 \;\;,  \label{gape}
\ee
which are the same as the classic BCS relation 
$E_k^2=\xi_k^2+\Delta_k^2$, modulo the dispersion renormalization 
factors $-A_1$ and $-C_3$ - c.f. eq. (\ref{dispe}).

\vspace{.3cm}

In case the dispersion is {\em not} given by eq. (\ref{gape}) above,
the determination of the {\em superfluid/superconducting gap} is less 
straightforward.
The gap in a physical system is probed through
various experimental techniques, and it is usually extracted from 
a fitting procedure to some {\em specific} theoretical models, including
{\em purely phenomenological ones}. E.g. for the high temperature 
superconductors these techniques include angle-resolved photoemission 
(ARPES), tunneling, NMR, Raman scattering, specific heat etc.
As far as the BCS theory is concerned,
things are pretty straightforward. Hence here one needs to calculate
the precise spectral response in terms of the microscopic parameters 
of $\Psi$ for any "gap"-probing 
experiment, and fit appropriately the data.

\vspace{.3cm}
$^*$ E-mail address : kast@iesl.forth.gr , giwkast@gmail.com


\begin{thebibliography}{99}
\bibitem{pse} Allowing for a continuous range of such momenta Q 
increases tremendously the computational complexity of the problem.
Hence one should at first look for a small discrete set of Q's.
\end{thebibliography}
\end{document}